\documentstyle[frontier,art10,psfig]{article}

\def\eg{\mbox{e.g.}}
\def\etal{\mbox{et~al.} }
\def\apgt{\ {\raise-.5ex\hbox{$\buildrel>\over\sim$}}\ }
\def\aplt{\ {\raise-.5ex\hbox{$\buildrel<\over\sim$}}\ }

\def\pe{\mbox{$P_{orb}$ -- $e$}~}

\newcommand{\pyr}{\mbox {{\rm yr$^{-1}$}}}

\newcommand{\msun}{\mbox{${\rm M}_\odot$}}
\newcommand{\myr}{ \mbox{ ${\rm M}_\odot \,{\rm yr}^{-1}$}}

\newcommand{\kms}{\mbox{${\rm km~s}^{-1}$}}

\newcommand{\ace}{\mbox {$\alpha_{\rm ce}$}}

\begin{document}
\heading{Evolution of  Close  Binaries:  Formation and Merger of \\
Neutron Star Binaries}
\par\medskip\noindent
\author{
Lev Yungelson$^{1,2}$, 
Simon F.\ Portegies Zwart$^{3,4,\!}$
\footnote[5]{Japan Society for the
Promotion of Science Fellow.
}
}
\address{
Institute of Astronomy of the RAS, 48 Pyatnitskaya Str., 109017 Moscow, Russia}
\address{
D.A.R.C., Observatoire de Paris, Section de Meudon, Meudon, France}
\address{Department of Information Science and Graphics, 
	 College of Arts and Science, 
	 University of Tokyo, 3-8-1 Komaba,
	 Meguro-ku, Tokyo 153, Japan }
\address{Astronomical Institute ``Anton Pannekoek'',
Kruislaan 403, NL-1098 SJ Amsterdam, The Netherlands
}

\begin{abstract}
We discuss the formation and evolution of binaries which contain 
neutron stars or black holes.
It is shown that in a
stellar system which for $10^{10}$\,yr had star formation rate
similar to the current one  in the Galactic disc, the rate of
neutron star mergers is $\sim 2 \times 10^{-5}$\,\pyr, consistent with
the observational estimates. The rate of black
hole and neutron star mergers is by an order of magnitude lower.
The pairs of black holes form but do not merge because heavy mass loss
by their progenitors efficiently moves the binary components apart.

\end{abstract}
\section{Introduction}
The wealth of observational features makes
the merging of close neutron star (NS) binaries one of
the most interesting phenomena in astrophysics.
Apart from the expected 
bursts of gravitational waves (GWR, \cite{CE77}),
$\gamma$-ray bursts \cite{BL84} and production of
$r$-process elements \cite{EI89} are possibly related processes.
Some NS binaries  are observed  as high mass binary pulsars (HMBP,
Table \ref{pulsars}, see \cite{PZY98} for references), but these are
possibly only a fraction of the entire population.
{\small
\begin{center}
\begin{tabular}[b]{lrrrcrrll}
\multicolumn{9}{l}{{\bf Table 1.} Observed population of high-mass
binary pulsars} \\
\hline \\
PSR      &  $P$    & $\log \dot{P}$&
$P_{orb}$ & e     & age & $\log B$ & $M_1$ & $M_2$ \\
         & [ms]   & [${\rm s\,s^{-1}}$]
&  [days]  &      & [Myr] & [G]  & [\msun] & [\msun] \\ \hline \\
J1518+49 &  40.94 &-19.4 &  8.634 & 0.249 &16000& 9.1 &
\multicolumn{2}{c}{${\rm M_t} = 2.62$} \\
B1534+12 &  37.90 &-17.6 &  0.420 & 0.274 &250 & 10.0   & 1.34 & 1.34\\
B1820-11 & 279.83 &-14.86&357.762 & 0.795 &3.3& 11.8 &
\multicolumn{2}{c}{$f(m)=0.7$} \\
B1913+16 &  59.03 &-17.1 &  0.323 & 0.617 &110 & 10.4 &  1.44 & 1.39 \\
B2127+11C&  30.53 &-17.3 &  0.335 & 0.681 &100 & 10.1 & 1.35 & 1.36 \\
B2303+46 & 1066.37 &-15.24& 12.340 & 0.658 &30& 11.9 & 
\multicolumn{2}{c}{${\rm M_t} = 2.60 $}\\ \hline \\
\label{pulsars}
\end{tabular}
\end{center}
}

All estimates for the Galactic merger rates of
NS binaries based on the observed
population arrive at values smaller than
$\simeq 10^{-5}$\,\pyr\ (\eg\ \cite{PH91}, \cite{HL96}).
Estimates based on simulations of the evolution of
close binaries arrive at rates up to
$\simeq 3 \times 10^{-4}$\,\pyr\ (\eg\ \cite{LPP97}, \cite{TY93}).
We will show that the estimated merger rates from observations and
model computations may be reconciled.

%
\begin{figure}[ht]
\centerline{\vbox{\psfig{file=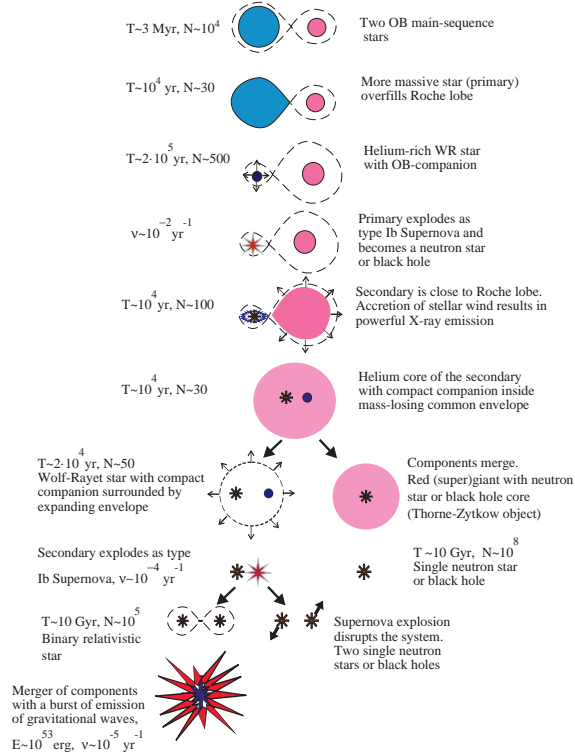,height=10cm}}}
\caption[]{Scenario of the evolution of massive close binaries.}
\label{figscen}
\end{figure}

\vspace{-12pt}
\section{Formation of neutron star binaries}

\subsection{Population synthesis}
``Theoretical'' estimates for the birth and merger rates of NS binaries
are usually derived from 
the  models which combine the
parameterized computations of
the evolution of stars with prescriptions for the
variation in the binary parameters due to the evolution of its
components.
By combining the model with distribution functions
for the zero-age primary masses, mass-ratios,
orbital periods and eccentricities, a complete
picture of the  population of binary stars is obtained.
An example of the evolutionary scenario for a binary, which
leads to the formation of two NS is given in
Fig. \ref{figscen}.

Although the concept of the formation of NS
binaries is relatively well understood 
(\cite{LGHC75}, \cite{TY73}, see however \cite{B95}),
the models imply  some badly
understood processes.
The most crucial among these are the
nonconservative mass exchange (especially the
common envelope phase) and the supernova process.
{\small
\begin{center}
\begin{tabular}[t]{lccc}
\multicolumn{4}{c}{{\bf Table 2.} Fraction of surviving systems  and relative
birthrates of NS binaries} \\ 
\hline\\
Event & No kick & Eq.(\ref{eqkick}) kick & Maxwellian kick\\
\hline\\
1st SN & 38\% & 4\% &  3\% \\
2nd SN & 10\% &  5\% &  4\% \\
${\rm \frac{NS+NS}{MS+MS}}$ &  4\% &  0.2\% & 0.1\% \\ [0.1cm]
${\rm \frac{NS+NS}{all~NS}}$ &  5\% &  0.2\% & 0.1\% \\[0.1cm]
\hline\\
\label{survival}
\end{tabular}

\end{center}
}

A common envelope (CE) forms when the accretor in a binary
is unable to acquire all matter lost by the donor.  The outcome of the
CE phase is either the coalescence of stars or the expulsion of the
envelope.  Models of the CE phase are still in an embryonic
state. Therefore, a  {\em common envelope parameter} \ace\ is usually
introduced. The most common definition of \ace\ is the ratio of the
binding energy of the CE to the orbital energy of the
binary (\eg\ \cite{IL93}).  Models of the populations of close
binary white dwarfs, low- and high-mass binary millisecond pulsars
(see below) and some other objects may be fit to observations if
$\ace \simeq 1 - 2.$ Though, formally, $\ace \apgt 1$~ means that
sources other than orbital energy must be invoked in the expulsion of
the CE, in fact, in such a parametric approach, this simply
suggests that energy deposited into a CE has to be
comparable to the orbital energy.  Roughly, the ratio of semi-major
axes of the orbit of a binary after and prior to the CE stage is
$a_f/a_i \propto \ace$, while the timescale for gravitational in-spiral 
is $\propto\,a_f^4$. This
makes all merger rate estimates highly sensitive to \ace. Estimates
given below were found by Portegies Zwart \& Yungelson
\cite{PZY98} for \ace=2.

A velocity kick may be imparted to the nascent NS.
While the mechanism and distribution of {\em natal
kicks} are still a
matter of debate, there is firm observational
evidence for their occurrence \cite{HP97}.
Natal kicks
may unbind binaries, which otherwise might have
remain bound and, {\em vice versa}, conserve the binaries which without 
the kick would have been dissociated.
An analysis of the role of kicks may be found \eg\ in
\cite{K96}, \cite{YSN93}. 

In \cite{PZY98} we used the distribution for isotropic kick velocities 
$v$ which has the functional form suggested in \cite{PA90} with
numerical parameters from \cite{H97}:
\begin{equation}
P(u)du = {4\over \pi} \cdot {du\over(1+u^2)^2},~~
u=v/\sigma,~~\sigma = 600~\kms.
\label{eqkick}
\end{equation}
Implementation of Eq.\,(\ref{eqkick}) 
provides the best fit to
the observed population of single pulsars close to the Sun
(\cite{HBWV97}, \cite{H97}). Peculiarity  of this distribution is
the combination of a peak at low velocity and of a high-velocity tail.
Selection effects in the population of single radio pulsars,
however, cause the exact shape of the kick velocity distribution to
be ill constrained both at low and high velocities. 
Alternatives as \eg\ a Maxwellian distribution with
mean of 250 to 500 \kms\ can be found in the recent literature
(\cite{HPH97}, \cite{LBH97}). 

\begin{figure}[h]
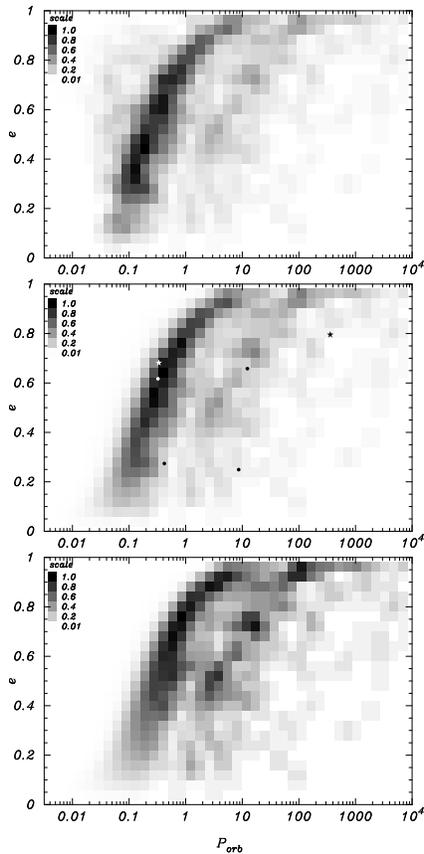

\centerline{\vbox{\psfig{file=a4q1pt0.ps,height=4cm,angle=-90}}}
\vspace*{-0.4cm}
\centerline{\vbox{\psfig{file=a4q1pt2.ps,height=4cm,angle=-90}}}
\vspace*{-0.4cm}
\centerline{\vbox{\psfig{file=a4q1pt3.ps,height=4cm,angle=-90}}}
\vspace*{-0.4cm}
\caption[]{
Probability distribution for orbital period and eccentricity for
the population of NS binaries
at the age $T = 0, 10^8~{\rm and}~10^9$\,yr.
The darkest shades correspond to the birthrate of $3.5 \times 10^{-7}\, \pyr$
(top panel) and numbers of 110 and 1100 (middle and bottom
panels, respectively).
Dots and asterisks show known HMBP.}
\label{pe}
\end{figure}

The results of computations without kicks, with
a kick taken isotropically from Eq. \ref{eqkick}, and with a kick
from a Maxwellian distribution with $\sigma = 450\,\kms$ are
presented in Table~2.   
Another parameters, like the initial mass-ratio
distribution or initial orbital
separation, affect the results by 
a factor of $\sim 2$.

\subsection{Merger rate of NS binaries in the Galactic disc}

A model which aims to predict a reliable birth and merger rate for
NS binaries   should provide
also the characteristics of related types of
binaries and HMBP.
For the HMBP, the most useful observational parameters
for comparison with the model are the orbital
period $P_{orb}$ and eccentricity $e$.

In \cite{PZY98} the best fit between
the observations of HMBP and the computations is obtained for
\ace=2 and the distribution of kick velocities given by
Eq. (\ref{eqkick}).
Figure \ref{pe} shows probability distributions of the orbital parameters
of the NS binaries with $P_{orb} \leq 10^4$\,day at an age of the
population of $T = 0, 10^8~ {\rm and}~ 10^9$\,yr. As demonstrated
in
\cite{PZY98}, natal kicks ``smear''  \pe\ distribution. In
the absence of kicks only PSR B1913+16 appears in
the region of the \pe\ diagram with high probability. Figure
\ref{pe} suggests that the average age of the population of observed HMBP
is several hundred Myr. Therefore it suggests that recycled pulsars are
born with low magnetic fields.

Table 3 compares the predictions of the preferred model from
\cite{PZY98} to the observations of NS binaries and related
objects. The model is normalized to
the current star formation rate in the Galactic disc of
$4\,\myr$~\cite{HJ97}; $M_{min} = 0.1\,\msun$
and 100\% binarity are assumed. The adopted lower mass limit for the
formation of a NS is 8\,\msun\ for an isolated
star and 11.4\,\msun\ for stars in close binaries.
The model satisfactorily reproduces the occurrence rates of
supernovae
descending from massive stars, numbers of Be/X-ray and massive X-ray
binaries (see also \cite{PZV96}). The model does not contradict
observational estimates of the birthrate of single
pulsars\footnote[6]{These estimates
strongly depends on the assumptions on the evolution of the
magnetic field, the distribution of the interstellar material and the
beaming fraction. For the latter $f = 0.3$~ was adopted.} or
the fraction of single recycled pulsars among all pulsars (see also
\cite{HPZV97}). It is
important to notice that our computations do not violate
the ``Bailes criterion'' \cite{B96}, which limits the fraction of HMBP
among the population of radio pulsars.

The birthrate of NS binaries
derived from computations is considerably
higher than the observational one, which involves
highly uncertain estimates of the incompleteness of the sample of
observed objects \cite{CL95} and their lifetimes \cite{HL96}.
{\em Observational} estimates are also likely to
underestimate the actual birthrate, since
a considerable fraction of old NSs in binaries may avoid
recycling and after the death of 
the young pulsar never show up as
pulsars\footnote[7]{Lipunov \etal\ \cite{LPP96} suggest that
recycled pulsars constitute only several per cent of the total
population of NS binaries; this estimate, however, 
weights heavily on the details of the accretion process on 
highly magnetized neutron stars.}.

\begin{figure}[h]
\centerline{\vbox{\psfig{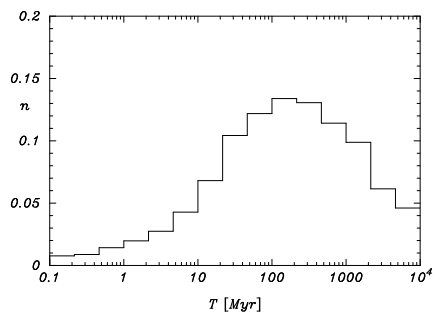}}}
\caption[]{Relative distribution of zero-age NS binaries over merger age.}
\label{age}
\end{figure}

The results of computations agree with observations for
the stages preceding formation of NS binaries,
give a reasonable estimate of their birthrate and 
reproduces their distribution in the \pe\ diagram.
We therefore expect that computations also correctly predict the
merger rate of NS binaries:
$2.3 \times 10^{-5}\,\pyr$~in the Galaxy. 
This estimate exceeds the most optimistic
``observational'' values by a factor of a few. However,
about 10\% to 20\% of all NS binaries merge only after the recycled
pulsar has died in 0.5 -- 1\,Gyr (Fig.~\ref{age}).
Another $\sim 10\%$~ merge within $\sim 10$ Myr, while the young
pulsar is still visible. And we note that possibly only a fraction of
old NS are recycled and show up in HMBP.
This suggests a revision of the ``observational'' merger
rate upward.
{\small
\begin{center}
\begin{tabular}{llll}
\multicolumn{4}{c}{{\bf Table 3.}
Descendants of massive binaries} \\
\hline \\
Object & Model & Observational estimate & Ref. \\
\hline\\
SNII+SNIb/c & $15\times 10^{-3}\,\pyr$ &  $(7 \div 21) \times
10^{-3}\,\pyr$ & \cite{CTT97}\\
Be/X-ray bin. & $10^3 - 10^4$ &  $\sim 2000$ \cite{MH89}\\
MXRB & $\sim 50$ &   $30 \div 80$ & \cite{ITY95}, \cite{MH89}\\
BH+MS/NS+MS & 8/60 & 4/30 & \cite{CKKS96}\\
Single PSR & $15\times 10^{-3}\,\pyr$ &  $(9\pm3)\times
10^{-3}\,\pyr$ & \cite{H97}\\
Single recycled NS/all NS & $\sim 1/100$ & $\sim 8/800$ & \cite{HPZV97}\\
Young NS+Old NS/all NS & $\sim 1/430$ & $ \aplt (1 \div 2)/800 $
& \cite{B96} \\
NS+BH/Single NS & $\sim 1/2000$ &  $<1/800$ \\
Birthrate of NS binaries & $3.4 \times 10^{-5} \pyr $ & $ 2.7 \times
10^{-5}
\pyr$ & \cite{CL95}\\
Merger rate NS+NS & $2\times10^{-5} \pyr$ & $0.8\times10^{-5}$
\pyr & \cite{HL96}
\\
Merger rate BH+NS & $0.1\times10^{-5} \pyr$ & ? \\
\hline \\
\end{tabular}
\label{theobs}
\end{center}
}

In \cite{PZY98} it is demonstrated
that a  higher average kick velocity in the supernova decreases the merger
rate. In the most unfavored case of a
Maxwellian kick combined with a mass ratio distribution
peaked to 0
and small \ace\ the merger rate 
may drop by an order of magnitude.

The lifetime for a NS binary before it merges is typically below 
1 to 2 Gyr (Fig.~\ref{age}).
The merger rate is therefore dominated by the star
 formation rate in the last few Gyr.
In massive ($\sim 10^{11}$\,
\msun) spiral (like the Milky Way) and irregular galaxies
the star formation rate in the far past was probably significantly
higher than at present \cite{S86}.
However, past 
bursts of star formation 
contribute only little to the
current merger rate of NS binaries.  The same point has to be  
made with respect to elliptical galaxies, as most of their stars were
formed in a short $(\aplt 1$\,Gyr) burst; the overwhelming majority of
NS binary mergers happened long ago.  
If we have two 10\, Gyr old, equal mass galaxies with the same average
star formation rate but in one case all stars are formed in a single
initial burst and in the other the star formation was constant,
the merger rate of NS binaries in
the first galaxy is $\sim 10$ to 15\% of that in the
latter.

Extrapolating our merger rates for NS binaries 
to the local Universe,   
using the blue luminosity of
the Universe \cite{PH91} and assuming $h=0.5$, we obtain for  
the detection rate by
the first generation of GWR detectors sensitive
to distances of $\aplt 25$\,Mpc \cite{BCT98}
only $\sim 1$ in 200 yr!
For advanced detectors (sensitive to
a distance of 250 Mpc) the
detection rate increases to $\sim 5$\,\pyr.

The merger rate for black hole + neutron star binaries (BH+NS)
is expected to be about 20 
times smaller than for NS binaries \cite{PZY98} and  
the rate of BH+BH mergers is virtually zero. 
The reason behind this is twofold.
First,
massive stars suffer heavy mass loss especially when
they become Wolf-Rayet
stars. E.g., evolutionary computations for solar composition stars with
{\em observationally inferred mass loss rates} \cite{SCH92} 
suggest that stars with an initial mass larger than $\sim 25$\,\msun\
have mass about 10\,\msun\ prior to supernova.
The stellar wind is expected to be isotropic and 
to carry away the specific angular momentum of the donor.
Therefore mass loss in a stellar wind increases the orbital separation.
Secondly, BH are not expected to receive sizable kicks which may,
if ``properly'' directed, decrease the binary separation. 

The signal-to-noise ratio for gravitational waves detectors is
given by \cite{BM94}:
\begin{equation} 
S/N \propto (M_1 M_2)^{1/2} (M_1+M_2)^{-1/6} D^{-1},
\label{sneq}
\end{equation}
where $M_{1,2}$\ are the masses of coalescing objects and $D$\ is the
distance to them. For $M_{BH} = 4 \div 17\, \msun$ suggested by
observed BH candidates, one would expect that a BH+NS merger 
is detectable to about 1.5 to 2.5 times as far as NS+NS
merger. Hence, the rates of detections of NS+NS and BH+NS mergers may
be comparable.

\subsection{Merger rate of NS binaries in globular
clusters}

Also in the dense cores of globular clusters (GC) mergers between 
NS may occur.
The fraction of primordial binaries which evolve to the stage where
GWR drives the components into coalescence
is, however, small. The majority of them are
ejected from the cluster or dissociated by the encounter with another
star. Some of the ejected binaries still result in a NS binary
and finally merge.
Single NS in the cores of high-density GC
may find another star and be captured by it either in a two- or three
body encounter. If two NS are able to form a binary in such 
a way this may increase the NS binary merger rate for post-collapse
clusters considerably.

A reliable estimate for the occurrence rate of such mergers is not
trivial as the evolution of the binaries and the stars are coupled in
a complex fashion. 
For solving this problem direct N-body simulations with a realistic
number of stars are required, for which the hardware simply does not
exist at present. 
Recently, however, Portegies~Zwart \etal\ \cite{PZ97a}, \cite{PZ97b}
developed a simplified model for simulating the interaction
between evolving stars and binaries while taking dynamical encounters
into account. The microscopic dynamics of close
encounters between single stars and binaries are
solved using semi-analytic and 3-body techniques, while the
macroscopic evolution of the stellar system is assumed to be
unaffected by its slowly changing stellar population.

The results of these computations
concerning the merging of NS binaries may be summarized as
follows: In the model for a core-collapsed cluster, the fraction of
merging primordial binaries is $\sim 60\%$.  The fraction of NS
binaries among these is $\sim 0.004.$
The time interval between core
collapse and Hubble time is $\sim 5$\,Gyr.
If one  takes a globular star cluster with
$10^6$ stars of which 10\% are in the region with a high density and
25\% primordial binaries, one obtains a rate for NS binary mergers of
$\sim 1.2 \times 10^{-8}\,\pyr$.

Of the binaries which are formed by tidal capture, which occurs once
every $10^7$\, years, the merger rate for NS binaries is only $4 \times
10^{-10}\, \pyr$ (almost all tidally formed binaries coalesce, but
only a fraction of 0.004 contains two NS).

For a total of $\sim 250$~GC in our Galaxy merger rate for NS binaries
becomes $\sim 10^{-7} \pyr$. The detection rate for the early GWR detectors
is therefore only $\aplt 1/1000$ years (depending on the fraction of dense
star clusters).  Thus, the rate of NS mergers in the field under any
circumstances is much higher than the rate of mergers in GC.

\section{Conclusion}
``Theoretical'' and ``observational'' merger rates of NS binaries produce 
similar results if the ejection of CE is highly
efficient and nascent neutron stars receive, on average, velocity 
kicks of the order of $(200 \div 300)$\,\kms.

The rate of NS mergers is $\sim 2 \times 10^{-5}\, \pyr$ for a galaxy
with constant astration rate of 4\myr\ (and $M_{min}=0.1\,
\msun$ and 100\% binarity). This rate is uncertain by a factor
of a few mainly due to uncertainties in the
kicks velocity distribution and the efficiency of energy
deposition into the CE.

Merger rate is mainly determined by the current star formation rate
and much less by the star formation history of the Galaxy.
The time to coalesce two NS  
stars is typically a few times $10^8$\,years.

Extrapolating the Galactic merger rate of NS binaries,
we arrive at a detection rate of once every $\sim 200$\,yr for 
the first generation GWR detectors (which are expected to be
sensitive up to $\sim 25$\,Mpc).
BH+NS mergers may be registered at the
rate comparable with binary NS mergers.

Mergers of BH+BH binaries are not to be expected because 
the severe mass loss in the stellar winds of their very massive 
progenitors results
in too large for merger orbital separations. 

Only $\sim 1\%$ of the GWR bursts are expected to originate
from GC.

\acknowledgements 
This work was supported by NWO Spinoza grant 08-0 to
E.~P.~J.~van den Heuvel, and RFBR grant 960216351.


\vfill
\end{document}